\documentclass[letterpaper]{article} % DO NOT CHANGE THIS
\usepackage{aaai2026}  % DO NOT CHANGE THIS
\usepackage{times}  % DO NOT CHANGE THIS
\usepackage{helvet}  % DO NOT CHANGE THIS
\usepackage{courier}  % DO NOT CHANGE THIS
\usepackage[hyphens]{url}  % DO NOT CHANGE THIS
\usepackage{graphicx} % DO NOT CHANGE THIS
\usepackage{natbib}  % DO NOT CHANGE THIS AND DO NOT ADD ANY OPTIONS TO IT
\usepackage{caption} % DO NOT CHANGE THIS AND DO NOT ADD ANY OPTIONS TO IT
\usepackage[ruled,vlined]{algorithm2e}  
\usepackage{multirow}
\usepackage{amsmath}
\usepackage{amsthm}
\usepackage{booktabs}

\usepackage{amsfonts}       % blackboard math symbols
\usepackage{subfigure}
\usepackage{cleveref}
\usepackage{array}

\title{Controllable Financial Market Generation with Diffusion Guided Meta Agent}
\author {
  Yu-Hao Huang\textsuperscript{\rm 1}, Chang Xu\textsuperscript{\rm 2}\thanks{Corresponding author.}, Yang Liu\textsuperscript{\rm 2}, Weiqing Liu\textsuperscript{\rm 2}, Wu-Jun Li\textsuperscript{\rm 1}, Jiang Bian\textsuperscript{\rm 2}\\
}
\affiliations {
    \textsuperscript{\rm 1}National Key Laboratory for Novel Software Technology, School of Computer Science, Nanjing University\\
    \textsuperscript{\rm 2}Microsoft Research Asia\\
    huangyh@smail.nju.edu.cn, \{chanx, yangliu2, weiqing.liu\}@microsoft.com,\\ 
    liwujun@nju.edu.cn,
    jiang.bian@microsoft.com
}

\begin{document}

\maketitle

\begin{abstract}
Generative modeling has transformed many fields, such as language and visual modeling, while its application in financial markets remains under-explored.
As the minimal unit within a financial market is an order, order-flow modeling represents a fundamental generative financial task.
However, current approaches often yield unsatisfactory fidelity in generating order flow, and their generation lacks controllability, thereby limiting their practical applications.
In this paper, we formulate the challenge of controllable
financial market generation, and propose a \textit{\underline{Di}ffusion \underline{G}uided \underline{M}eta \underline{A}gent}~(DigMA) model
to address it. Specifically, we employ a conditional diffusion model to capture the dynamics of the market state represented by time-evolving distribution parameters of the mid-price return rate and the order arrival rate, and we define a meta agent with financial economic priors to generate orders from the corresponding distributions. 
Extensive experimental results show that DigMA achieves superior controllability and generation fidelity. 
Moreover, we validate its effectiveness as a generative environment for downstream high-frequency trading tasks and its computational efficiency.
\end{abstract}

\begin{links}
    \link{Code}{https://github.com/microsoft/TimeCraft}
    \link{Extended version}{https://arxiv.org/abs/2408.12991}
\end{links}

\section{Introduction}\label{sec:intro}
Generative modeling has transformed fields such as natural language processing~\cite{gpt3,chowdhery23,singhal23}, media synthesis~\cite{rombach22,lu24,copet24}, science discovery~\cite{wang21,skinnider21,m24,furrutter24} and medical applications~\cite{marco23,tu24}, with techniques such as Generative Adversarial Networks~(GANs)~\cite{goodfellow2014generative}, Diffusion Models~(DMs)~\cite{ho20}, and pre-trained Transformers~\cite{vaswani2017attention} serving as core building blocks.
While the financial market is a data-intensive and intricate system characterized by dynamic interactions among participants and has attracted research attention for decades, the application of generative models into this domain remains rare. 
Similar to words in language and pixels in images, an order is the fundamental element representing the minimal unit of events within a financial market~\cite{palmer94,lux99,raberto01,chiarella02,chiarella09}.
Through sequential order flow, researchers, investors, and policymakers can investigate the intrinsic interactive mechanisms and microstructure of financial markets~\cite{raberto01,takanobu20,coletta21,quant4}.
Therefore, order-flow modeling constitutes a fundamental generative task in financial domains.

Recent works have attempted to simulate order-level financial markets using agent-based methods, employing either rule-based agents~\cite{vyetrenko20,abides20,abides21} or learned agents~\cite{li20,coletta23,li2025mars}. 
However, their fidelity and flexibility remain limited.
Ruled-based agents rely on over-simplified assumptions about the market and can only represent known types of market participants under predefined scenario. They are not trained on real market data but are instead constructed using hand-crafted rules, resulting in limited simulation fidelity.
Learned agents are trained to predict next order given history order flow, where the order flow may contain hundreds of orders within a single minute. It is challenging to directly capture the long term dependency, as a trading day spans hundreds of minutes. These models tend to emphasize the local distribution of the simulated order flow while neglecting the global dynamic.
More importantly, controllability of the generated market, which would enable researchers and practitioners to systematically explore market behaviors under various conditions such as extreme or rare events, is absent in the literature. This underscores a practical gap for conducting scenario-based experiments and counterfactual analysis~\cite{takanobu20,quant4}.

In contrast to existing works, we propose incorporating controllability into financial market modeling by formulating the problem as a conditional generation task.
The objective is to construct specific scenarios characterized by varying levels of asset return, intraday volatility, and rare events such as sharp drops or extreme amplitudes. 
Achieving this requires addressing the critical challenge of establishing a principled connection between high-level control targets, such as desired market scenarios, and the orders generated by the model.
Moreover, exercising control at the order level is particularly challenging, because (1) the long and irregular length of real-world order-flow sequences hinders the direct application of diffusion models to raw order-level data, and (2) linking a ``macro'' control target to each individual ``micro'' order is impractical due to the inherently low signal-to-noise ratio present in order flow.

In this paper, we formulate the challenge of controllable financial market generation, and propose the 
\textit{\underline{Di}ffusion \underline{G}uided \underline{M}eta \underline{A}gent}~(DigMA) 
model to address the challenge. 
Specifically, we employ a conditional diffusion model to capture the dynamics of the market state represented by time-evolving distribution parameters of the mid-price return rate and the order arrival rate, and we define a meta agent with financial economic priors to generate orders from the corresponding distributions.
With DigMA, we are able to control the generation process to simulate order flow under target scenario with high fidelity. Our contributions can be summarized as follows:

\begin{itemize}
    \item We formulate controllable financial market generation problem as a novel and pratically meaningful challenge for machine learning application in finance.
    
    \item To the best of our knowledge, 
    DigMA is among the pioneering models to integrate advanced diffusion-based generative techniques into financial market modeling.
    
    \item Through extensive experiments on real stock market data, we demonstrate that DigMA achieves superior controllability and the highest fidelity to established stylized facts. Moreover, we validate its effectiveness as a generative environment for downstream high-frequency trading tasks, as well as its computational efficiency.

\end{itemize}

\section{Preliminaries and Related Work}\label{sec:rw}
In this section, we briefly introduce preliminaries of the limit order book and review related research on financial market simulation and diffusion models.

\subsection{Limit Order Book}
The majority of financial markets worldwide operate under a double-auction mechanism, in which orders serve as the fundamental units of events.
An order consists of four basic elements: timestamp $t$, price $p$, quantity $q$, and order type $o$. Real markets feature a variety of order types, such as limit orders, market orders, cancel orders, and conditional orders. In the literature, it is often sufficient to represent trading decisions using limit orders and cancel orders~\cite{chiarella09}.

\begin{figure}
  \centering
    \addtolength\abovecaptionskip{-0.3cm}
    \addtolength\belowcaptionskip{-0.7cm}
  
\includegraphics[width=\linewidth]{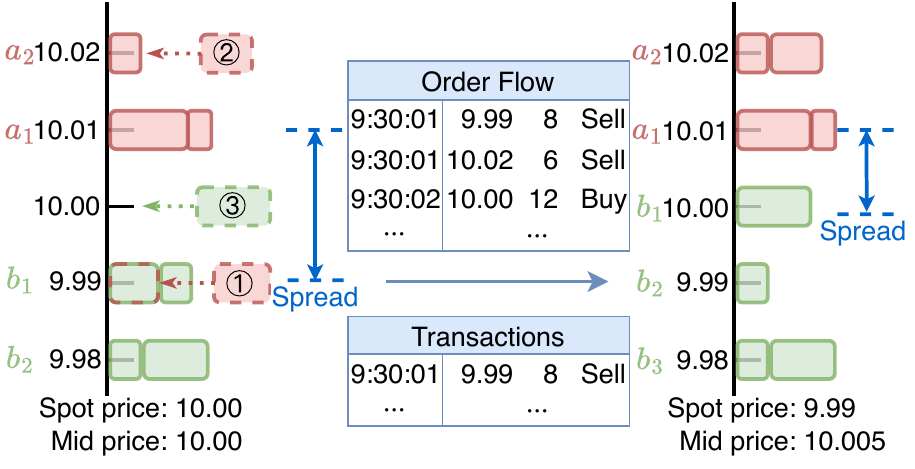}
  \caption{Limit order book and order flow}
  \label{fig:lob}
\end{figure}

The output of a market simulation model is a sequence of orders $\boldsymbol{O} = \{(t_1,p_1,q_1,o_1), (t_2,p_2,q_2,o_2), \ldots\}$, commonly referred to as the order flow, which collectively determines the order book and the resulting price series.
The order book is defined as the set of outstanding orders that have not yet been executed. Limit orders can be further categorized into buy and sell limit orders, whose prices are referred to as bids and asks, respectively. An order book consisting exclusively of limit orders is known as the limit order book~(LOB). The transaction price at each timestamp is typically defined as the price of the most recently executed order. Sampling these prices at a chosen frequency yields the corresponding price series.
An illustrative example of a limit order book is shown in Figure~\ref{fig:lob}.

\subsection{Financial Market Simulation}
Early work on financial market simulation adopted a multi-agent approach~\cite{raberto01,palmer94,lux99}, using rule-based agents under simplified trading protocols to replicate ``stylized facts''~\cite{cont01}, such as volatility clustering.
\citet{chiarella02,chiarella09} extended these simulations to order-driven markets, aligning more closely with modern stock market structures. Subsequent studies further customized agent behaviors within this framework to better support research on decision-making~\cite{vyetrenko20,abides20,wang17}.
With recent advances in machine learning, researchers have also explored neural networks as world agents capable of directly predicting orders from historical order flow~\cite{coletta21,li20,coletta22,coletta23}.
In contrast, our model employs a conditional diffusion model to guide an agent-based system in order generation.

\subsection{Diffusion Models}
Diffusion probabilistic models, commonly referred to as diffusion models, learn to transform between simple and target distributions through a sequence of small perturbations defined by a diffusion process~\cite{SD15,ho20,song21}. These models have been applied to data generation across a wide range of modalities, including images~\cite{rombach22,marco23,dharwial21,song23}, audio~\cite{kong21,liu2023audioldm}, video~\cite{sora,guo2024animatediff}, and general time series~\cite{tsdiff,mgtsd,diffts,huang2025timedp,bridge,tardiff}.
In contrast to existing work, we are the first to apply diffusion models to the generation of financial orders.

\section{Method}\label{sec:mth}

\begin{figure*}[!t]
  \centering
    \addtolength\abovecaptionskip{-0.4cm}
    \addtolength\belowcaptionskip{-0.4cm}
  \subfigure[Meta controller and order generator in DigMA.]{
  \label{fig:overview_a}
  \includegraphics[width=0.48\textwidth]{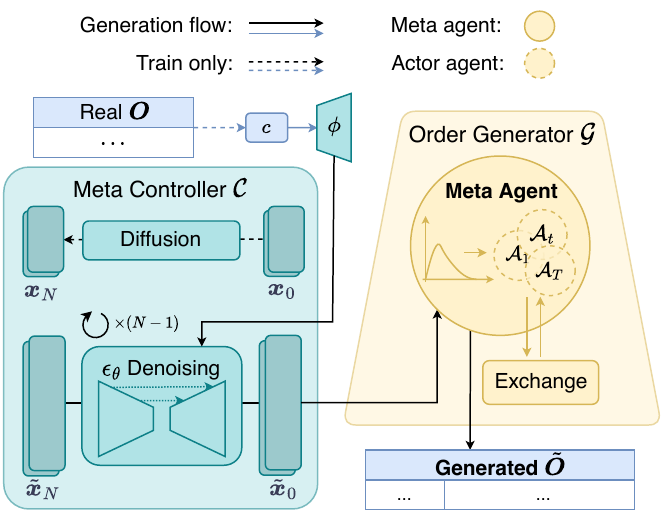}
  }
  \subfigure[Data flow of DigMA.]{
  \includegraphics[width=0.48\textwidth]{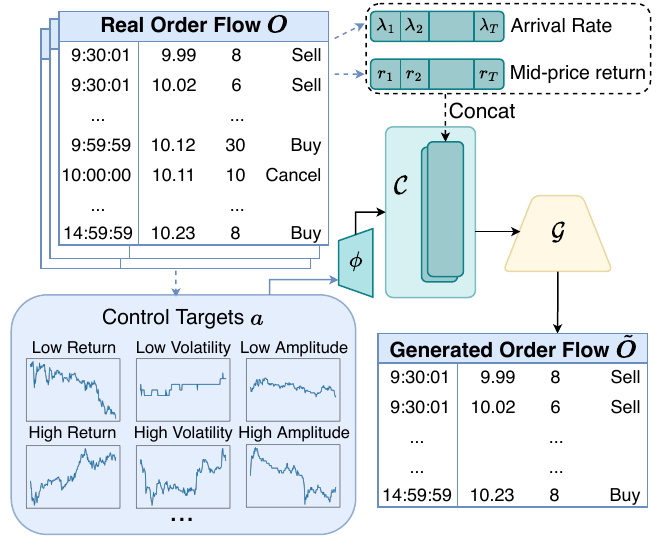}
  }
  \caption{Overview of DigMA model. Raw order flow is proccessed into market states for the meta controller to learn. The meta agent is guided by the meta controller, and generates simulated order flow.}
  \label{fig:overview}
\end{figure*}
In this section, we first present the problem formulation for controllable financial market generation. We then describe the architecture of our \textit{\underline{Di}ffusion \underline{G}uided \underline{M}eta \underline{A}gent}~(DigMA) model in detail.

\subsection{Problem Formulation: Controllable Financial Market Generation}

Unlike standard market simulation, which generates order flow unconditionally, controllable market generation aims to simulate order flows corresponding to specified scenarios using generative models. Such scenarios can be characterized by indicators representing aggregated order-flow statistics, including daily return, daily amplitude, and intraday volatility. Formally, let $\mathcal{F}$ denote a function that computes a given indicator. For a real order-flow sample $\boldsymbol{O} \sim q(\boldsymbol{O})$, the corresponding indicator value $a$ is computed as $a=\mathcal{F}(\boldsymbol{O})$.

In controllable financial market generation, a market generator $\mathcal{M}$ defines a conditional sampler $p_\mathcal{M}(\boldsymbol{O}\mid a)$. Given a control target $a$, the generator produces an order-flow sample $\tilde{\boldsymbol{O}} \sim p_\mathcal{M}(\boldsymbol{O}\mid a)$. The controllability objective is to minimize the discrepancy between the target $a$ and the indicator value of the generated order flow, $\tilde{a}=\mathcal{F}(\tilde{\boldsymbol{O}})$. Formally, the controllability objective is given by
\begin{equation}
    \min_{\mathcal{M}} \mathbb{E}_{a,\tilde{a}}\!\left[\lVert \tilde{a}-a \rVert^2\right]
    = \min_{\mathcal{M}} \mathbb{E}_{a,\boldsymbol{\tilde{O}}}\left[\lVert \mathcal{F}(\boldsymbol{\tilde{O}})-a \rVert^2\right].
\end{equation}

In addition, the generator is required to produce order flows with high fidelity. This objective aligns with the general goal of market simulation, which is to minimize a divergence measure $\mathcal{D}$ between the distributions of ``stylized facts'' computed from real and generated order flows. Here, the stylized facts can also be expressed as a set of aggregated statistics with corresponding computation functions. Let $\mathcal{F}'$ denote a function that computes a stylized fact, and let $p(\cdot)$ denote a probability density. The fidelity objective is then defined as
\begin{equation} 
\min_\mathcal{M} \mathcal{D}\left(p(\mathcal{F'}(\Tilde{\boldsymbol{O}}))\parallel p(\mathcal{F'}(\boldsymbol{O}))\right). \end{equation}

\subsection{Diffusion Guided Meta Agent Model}
Order flow data are highly intricate and noisy, often consisting of tens of thousands of events per stock daily, which poses significant computational challenges. Modeling the full distribution of order flow across diverse scenarios is therefore non-trivial. Moreover, directly linking a ``macro'' control target to individual ``micro'' orders is impractical due to the low signal-to-noise ratio of order flow. To address these challenges, we adopt a two-stage design instead of applying diffusion models directly to raw order flow.

DigMA consists of two modules. The first is a meta controller $\mathcal{C}$, which learns the intraday dynamics of latent market states $\boldsymbol{x}$ conditioned on a scenario $c$, and models the distribution $q(\boldsymbol{x}\mid c)$ using a conditional diffusion model. The second is an order generator $\mathcal{G}$, which comprises a simulated exchange and a meta agent. The meta agent incorporates financial economic priors and is guided by the meta controller to generate orders through a stochastic process.
An overview of the DigMA architecture is shown in~\Cref{fig:overview}. Overall, the model is expressed as $\mathcal{M} = \{\mathcal{C},\mathcal{G}\}$.

\subsubsection{Meta Controller}
To represent intraday market dynamics associated with a given scenario $c$, the market states $\boldsymbol{x}$ are required to evolve over time while maintaining a causal relationship with  $c$.
We therefore extract the minutely mid-price return rate $\boldsymbol{r}$ and the order arrival rate $\boldsymbol{\lambda}$ from real order-flow data, and define the market states as $\boldsymbol{x} = \{\boldsymbol{r}, \boldsymbol{\lambda}\}$.

Since the number of trading minutes within each trading day is fixed, the stacked market-state sequence corresponding to a single day naturally forms one training sample.
The objective is to fit the distribution of these market states:
\begin{equation}
    \min \; \mathbb{E}_{\boldsymbol{x}}\mathcal{D}\!\left(p_\mathcal{C}(\boldsymbol{x}) \parallel q(\boldsymbol{x})\right).
\end{equation}

Given training samples $\{\boldsymbol{x} \sim q(\boldsymbol{x})\}$, we generate noisy latent variables as
$\boldsymbol{x}_{n}=\sqrt{\bar{\alpha}_{n}} \boldsymbol{x}_{0}+\sqrt{1-\bar{\alpha}_n} \boldsymbol{\epsilon}$,
where $\boldsymbol{\epsilon} \sim \mathcal{N}(\boldsymbol{0}, \boldsymbol{I})$ and $n$ denotes the diffusion step.
Here, $\bar{\alpha}_n$ is derived from the variance schedule $\{\beta_n\in(0,1)\}_{n=1}^{N}$, where $\alpha_n=1-\beta_n$ and $\bar{\alpha}_n = \prod_{i=1}^n\alpha_i$, and $N$ denotes the maximum diffusion step.
We adopt the $\boldsymbol{\epsilon}$-parameterized denoising diffusion probabilistic model~(DDPM)~\cite{ho20} and train it to predict the injected noise from $\boldsymbol{x}_n$ as $\hat{\boldsymbol{\epsilon}}=\boldsymbol{\epsilon}_\theta(\boldsymbol{x}_n,n)$, where $\theta$ denotes the model parameters.
The resulting training loss is defined as
\begin{equation}
    L_M := \mathbb{E}_{\boldsymbol{x},\boldsymbol{\epsilon}\sim \mathcal{N}(\boldsymbol{0},\boldsymbol{I}),n}
    \big[\|\boldsymbol{\epsilon}-\boldsymbol{\epsilon}_\theta(\boldsymbol{x}_n,n)\|^2\big].
\end{equation}

During sampling, the diffusion process proceeds iteratively from $\boldsymbol{x}_N$ to $\tilde{\boldsymbol{x}}_0$ following
\begin{equation}\label{eq:ddpmsample}
    \tilde{\boldsymbol{x}}_{n-1}=\frac{1}{\sqrt{\alpha_n}}
    \Big(\tilde{\boldsymbol{x}}_n-\frac{1-\alpha_n}{\sqrt{1-\bar{\alpha}_n}}
    \boldsymbol{\epsilon}_\theta(\tilde{\boldsymbol{x}}_n,n)\Big)+\sigma_n\boldsymbol{z},
\end{equation}
where $\boldsymbol{x}_N \sim \mathcal{N}(\boldsymbol{0},\boldsymbol{I})$, $\boldsymbol{z} \sim \mathcal{N}(\boldsymbol{0}, \boldsymbol{I})$, and $\sigma_n=\sqrt{\beta_n}$.
The resulting $\tilde{\boldsymbol{x}}_0$ provides a latent market trajectory that enables the meta controller to guide the meta agent during the subsequent order generation process.

To further enable controlled order generation, we condition the diffusion model on target scenario variables so as to steer the generated order flow toward the desired market regime.
Following common practice~\cite{rombach22}, we implement a conditional $\boldsymbol{\epsilon}$-parameterized noise predictor $\boldsymbol{\epsilon}_\theta(\boldsymbol{x}_n,n,c)$ to support sampling under a specified control target $c$.
Specifically, we adopt three indicators that are commonly used to characterize financial market states as control targets: daily return, amplitude, and volatility.
Each of these indicators can be computed from the return rates of the price series, and controlling them allows the generated order flow to reflect a wide range of realistic market scenarios.

To incorporate control targets into the diffusion model, we introduce a target-specific feature extractor $\boldsymbol{\phi}$ that projects target indicators into latent representations $\boldsymbol{\phi}(\boldsymbol{c})$.
We design two types of control encoders.
The first is a \textit{discrete} control encoder, where target conditions are mapped into predefined bins that are treated as class labels and embedded via a learnable embedding matrix.
The second is a \textit{continuous} control encoder, which uses a fully connected network to map real-valued conditions into latent vectors.
The feature extractor $\boldsymbol{\phi}$ is trained jointly with the conditional sampler using
\begin{equation}
    L_C := \mathbb{E}_{\boldsymbol{x},\boldsymbol{c},\boldsymbol{\epsilon}\sim \mathcal{N}(\boldsymbol{0},\boldsymbol{I}),n}
    \big[\|\boldsymbol{\epsilon}-\boldsymbol{\epsilon}_\theta(\boldsymbol{x}_n,n,\boldsymbol{\phi}(\boldsymbol{c}))\|^2\big].
\end{equation}

For both encoders, we employ classifier-free guidance~\cite{ho21} to perform control.
During training, unconditional and conditional samplers are jointly optimized by randomly dropping the conditioning information.
During sampling, the guided score is computed as
\begin{equation}
    \tilde{\boldsymbol{\epsilon}}_{\theta,\phi}(\boldsymbol{x}_n,n,\boldsymbol{c})
    =(1-s)\boldsymbol{\epsilon}_\theta(\boldsymbol{x}_n,n)
    +s\boldsymbol{\epsilon}_\theta(\boldsymbol{x}_n,n,\boldsymbol{\phi}(\boldsymbol{c})),
\end{equation}
where $s$ is a scaling factor that controls the guidance strength.
Sampling then follows Equation~\ref{eq:ddpmsample} with
\begin{equation}
    \boldsymbol{x}_{n-1}=\frac{1}{\sqrt{\alpha_n}}
    \Big(\boldsymbol{x}_n-\frac{1-\alpha_n}{\sqrt{1-\bar{\alpha}_n}}
    \tilde{\boldsymbol{\epsilon}}_{\theta,\phi}(\boldsymbol{x}_n,n,\boldsymbol{c})\Big)+\sigma_n\boldsymbol{z}.
\end{equation}
In practice, we adopt DDIM sampling~\cite{ddim} to improve sampling efficiency.
For the model backbone, we use a U-Net constructed primarily from 1D convolutional layers, with parameters shared across diffusion time steps.
Additional implementation details are provided in the appendix.

\begin{figure*}[!t]

  \addtolength\abovecaptionskip{-0.2cm}
  \centering
  \includegraphics[width=0.9\textwidth]{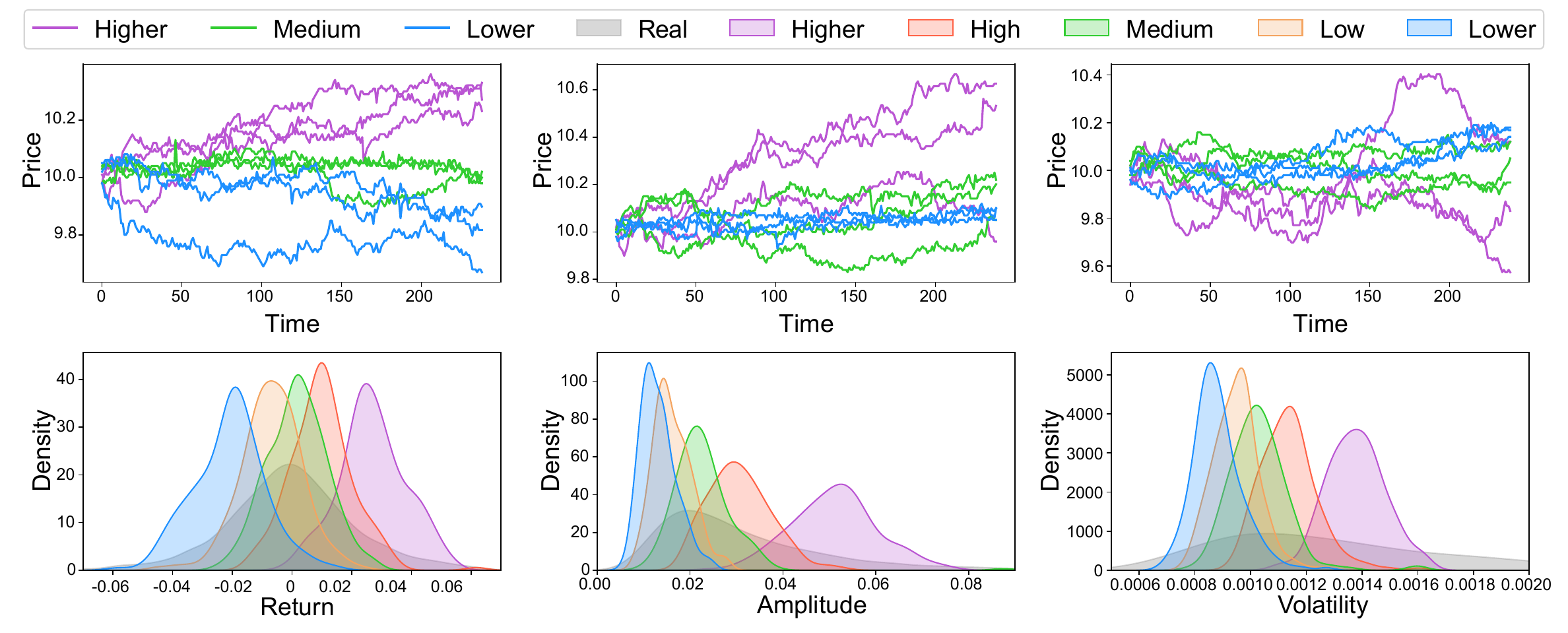}
  
  \caption{Aggregated price curves~(first row) and distributions of targeted indicators~(second row). In the first row, each curve represents the price trajectory derived from one day of generated order flow. In the second row, each colored density corresponds to the distribution of a targeted indicator computed from the generation results.}
  \label{fig:ctrl}
\end{figure*}

\begin{table*}[!t]
\small
\centering
\addtolength\belowcaptionskip{-0.3cm}
\begin{tabular}{@{}cccccccccccc@{}}
\toprule
 &  & \multicolumn{5}{c}{\textbf{A-Main}} & \multicolumn{5}{c}{\textbf{ChiNext}} \\ \midrule
\textbf{Target} & \textbf{Method} & \textbf{Lower} & \textbf{Low} & \textbf{Medium} & \textbf{High} & \textbf{Higher} & \textbf{Lower} & \textbf{Low} & \textbf{Medium} & \textbf{High} & \textbf{Higher} \\ \midrule
\multirow{3}{*}{Return} & No Control & 1.443 & 0.583 & 0.529 & 0.813 & 2.337 & 0.979 & 0.684 & 0.992 & 1.718 & 3.923 \\
 & Discrete & 1.055 & 0.494 & 0.228 & 0.429 & 0.664 & 1.285 & 0.807 & 0.243 & \textbf{0.413} & 0.869 \\
 & Continuous & \textbf{0.206} & \textbf{0.178} & \textbf{0.161} & \textbf{0.184} & \textbf{0.212} & \textbf{0.584} & \textbf{0.539} & \textbf{0.342} & 0.449 & \textbf{0.840} \\ \midrule
\multirow{3}{*}{Amplitude} & No Control & 0.521 & 0.268 & 0.268 & 0.699 & 3.298 & 1.130 & 0.638 & 0.427 & 0.608 & 2.763 \\
 & Discrete & \textbf{0.049} & 0.088 & 0.309 & 0.502 & 0.930 & \textbf{0.057} & 0.134 & 0.346 & 0.523 & \textbf{0.963} \\
 & Continuous & 0.054 & \textbf{0.076} & \textbf{0.149} & \textbf{0.247} & \textbf{0.348} & 0.110 & \textbf{0.116} & \textbf{0.255} & \textbf{0.437} & 0.973 \\ \midrule
\multirow{3}{*}{Volatility} & No Control & 0.021 & 0.115 & 0.431 & 1.209 & 4.288 & 0.029 & 0.246 & 0.713 & 1.737 & 5.221 \\
 & Discrete & 0.016 & 0.123 & 0.383 & 0.890 & 2.393 & 0.029 & 0.188 & 0.481 & \textbf{0.948} & \textbf{2.257} \\
 & Continuous & \textbf{0.011} & \textbf{0.104} & \textbf{0.318} & \textbf{0.774} & \textbf{2.389} & \textbf{0.028} & \textbf{0.178} & \textbf{0.473} & 1.016 & 2.631 \\ \bottomrule
\end{tabular}
\caption{Mean squared error~(MSE) between the targeted indicators and the aggregated statistics computed from the generated order flow. Best results are highlighted in bold.}
\label{tab:ctrl}
\end{table*}

\subsubsection{Order Generator}
The order generator consists of a simulated exchange and a meta agent. The simulated exchange replicates the double-auction market protocol on which the majority of financial markets operate. It facilitates agent--market interactions and provides the foundation for producing realistic order-level market dynamics.
The meta agent represents the aggregate behavior of all traders in the generated market, functioning as a world agent. Unlike prior work that employs learned agents as world agents, our meta agent incorporates financial economic priors and is guided by the meta controller.

Specifically, the meta agent generates orders through a stochastic process, whose key parameters are supplied by the meta controller.
For each trading minute $t$, the meta agent ``wakes up'' after a time interval $\delta_i$ drawn from an exponential distribution $f(\delta_i;\lambda_t)=
\lambda_t e^{-\lambda_t \delta_i}$, where $i$ indexes the total number of wake-ups within a trading day and $\lambda_t$ is the arrival rate provided by the meta controller.
Upon each wake-up, the meta agent instantiates an actor agent $\mathcal{A}_i$ from a family of heterogeneous agents~\cite{chiarella09}. The agent makes decisions by optimizing a CARA utility function based on current market observations. The order-generation procedure proceeds as follows:
\begin{itemize}
\item \textbf{Initialization.} The actor agent is initialized with a random holding position $S$, the corresponding cash balance $C$, and random component weights $g_f, g_c, g_n$ for its three heterogeneous components, namely fundamental, chartist, and noise. These weights are drawn independently from exponential distributions whose expected values follow a ratio of $10{:}1.5{:}1$.
\item \textbf{Return estimation.} The actor agent estimates the objective future return $\hat{r}$ as a weighted average of the fundamental, chartist, and noise components. The fundamental component is given by the return $r_t$ determined by the meta controller, the chartist component is the historical average return $\bar{r}$ obtained from the simulated exchange, and the noise component is a small Gaussian perturbation $r_\sigma$. Accordingly,
$\hat{r}=\frac{g_f r_t+g_c\bar{r}+g_n r_\sigma}{g_f+g_c+g_n}$.
\item \textbf{Holding optimization.} Given the estimated return, the actor agent computes a future price estimate $\hat{p}_t=p_t\exp{(\hat{r})}$. By optimizing the CARA utility over future wealth~\cite{chiarella09}, the actor derives its demand function
$u(p)=\frac{\ln{(\hat{p}_t/p)}}{aVp}$,
where $a$ is the risk-aversion coefficient and $V$ denotes historical price volatility. The actor then identifies the lowest acceptable price $p_l$ satisfying $p_l(u(p_l)-S)=C$, ensuring affordability under the current cash balance.
\item \textbf{Order sampling.} The actor agent samples an order price uniformly between the lowest acceptable price and the estimated price, $p_i\sim \mathcal{U}(p_l,\hat{p}_t)$. It then determines the order volume as $q_i=u(p_i)-S$ and the order type as $o_i=\operatorname{sign}(q_i)$, where $o_i=1$ indicates a buy order and $o_i=0$ indicates a sell order.
\end{itemize}
The resulting order is recorded as $\boldsymbol{o}_i = (t_i,p_i,q_i,o_i) \sim p(\boldsymbol{o}\mid r_t,\lambda_t)$,
where $t_i = \sum_{j=1}^i{\delta_j}$. Pseudocode for the order-generation procedure is provided in Algorithm~2 in the appendix.

Order generation terminates at $t_{\max}$ when the next wake-up time $t_i$ would exceed the trading hours of the day. The final generated order flow is recorded as
\begin{equation}
    \tilde{\boldsymbol{O}} = \{\boldsymbol{o}_1,\ldots,\boldsymbol{o}_{\max}\} \sim p(\boldsymbol{O}\mid\boldsymbol{\tilde{x}}),
\end{equation}
where $\boldsymbol{\tilde{x}} = \{\boldsymbol{r},\boldsymbol{\lambda}\}$ is generated by the meta controller.

\section{Experiments}\label{sec:exp}
In this section, we describe the experimental setup and present results on real-world datasets to evaluate both the controllability and fidelity of DigMA. We further demonstrate the usefulness of DigMA as a generative environment for a high-frequency trading reinforcement learning task and analyze its computational efficiency.

\subsection{Dataset and Model Configurations}
We conduct experiments on two tick-by-tick order-flow datasets from global markets: \textit{A-Main} and \textit{ChiNext} from the Chinese stock market. For each dataset, we use 5,000 samples for validation and 5,000 samples for testing, with the remaining samples used for training. Additional details on dataset preprocessing, along with links to the code and illustrative examples, are provided in the appendix.

We train the diffusion model on each dataset for 10 epochs with 200 diffusion steps using the AdamW optimizer. The mini-batch size is set to 256, and the learning rate is $1\times 10^{-5}$. For both discrete and continuous control models, conditioning information is randomly dropped with probability 0.5 during training. Other implementation details are provided in the appendix.

\begin{figure*}[!t]
  \centering
\addtolength\abovecaptionskip{-0.2cm}
\addtolength\belowcaptionskip{-0.2cm}
  \includegraphics[width=0.85\textwidth]{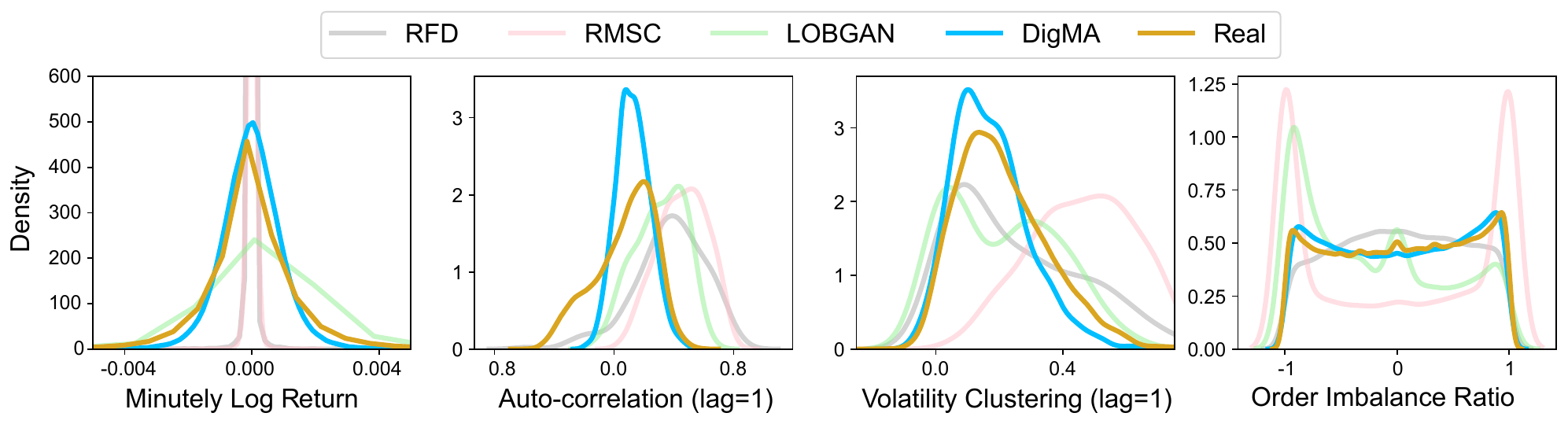}
  \caption{Comparison of stylized facts distribution across baselines. The x-axis is the stylized facts and the y-axis is the density.}
  \label{fig:sfd}
\end{figure*}

\begin{table*}[!t]
\small
\centering
  \addtolength\belowcaptionskip{-0.5cm}
\begin{tabular}{@{}lrrrrrrrr@{}}
\toprule
 & \multicolumn{4}{c}{\textbf{A-Main}} & \multicolumn{4}{c}{\textbf{ChiNext}} \\ \midrule
\textbf{Model} & \textbf{MinR} & \textbf{RetAC} & \textbf{VolC} & \textbf{OIR} & \textbf{MinR} & \textbf{RetAC} & \textbf{VolC} & \textbf{OIR} \\ \midrule
RFD & 1.198 & 5.010 & 0.839 & 0.015 & 0.272 & 2.987 & 0.691 & 0.022 \\
RMSC & 2.640 & 10.170 & 1.237 & 0.563 & 1.371 & 7.461 & 0.668 & 0.588 \\
LOBGAN & 0.151 & \textbf{1.903} & 1.101 & 0.309 & 0.135 & \textbf{1.711} & 0.507 & 0.282 \\
DigMA & \textbf{0.084} & 2.781 & \textbf{0.273} & \textbf{0.009} & \textbf{0.079} & 1.997 & \textbf{0.218} & \textbf{0.009} \\ \bottomrule
\end{tabular}
\caption{K-L divergence of stylized facts distribution between real and simulated order flow.}
\label{tab:dist}
\end{table*}

\subsection{Evaluation on Controlling Financial Market Generation}
In this experiment, we evaluate DigMA’s ability to perform controllable market generation. To enable conditioning on target scenarios, we train DigMA using three indicators: \textit{return}, \textit{amplitude}, and \textit{volatility}, each capturing a broad range of market conditions. For each indicator, we first obtain its empirical distribution from the real-world order-flow dataset. We then partition the values into five percentile-based bins, representing \textit{lower}, \textit{low}, \textit{mid}, \textit{high}, and \textit{higher} value cases for the scenario characterized by the corresponding indicator. DigMA with a discrete control encoder is trained using these bin labels as categorical conditions. For the continuous control encoder, we use normalized numerical indicator values as inputs. During testing, discrete models take the corresponding class labels as control targets, while continuous models use the median indicator value of each bin computed from real samples.

Table~\ref{tab:ctrl} reports the \textbf{mean squared error~(MSE)} between each control target (i.e., bin median) and the corresponding indicator computed from the generated order flow, for DigMA with both discrete and continuous control encoders. Results are averaged over three runs with different random seeds. The ``No Control'' baseline is a DigMA variant in which the conditioning mechanism of the meta-controller is removed, resulting in generation that is independent of the target scenario. As shown in~\Cref{tab:ctrl}, DigMA achieves consistently low errors across all indicators, whereas the uncontrolled baseline produces either random or scenario-insensitive outputs.

Figure~\ref{fig:ctrl} visualizes the mid-price series for controlled generation samples across different scenarios, together with the distributions of indicators computed from 200 independent runs per scenario. The price trajectories reflect the intended market conditions, and the indicator distributions shift appropriately in response to the control targets. These results demonstrate that DigMA effectively enables controllable financial market generation.

\subsection{Evaluation on Generation Fidelity}
We evaluate the generation fidelity of DigMA and compare it with both rule-based and learning-based baselines:
\begin{itemize}
    \item \textbf{RFD}~\cite{vyetrenko20} is a market simulation configuration featuring heterogeneous agents, including 1 market maker, 25 momentum traders, 100 value traders, and 5,000 noise traders with random fundamentals.
    \item \textbf{RMSC}~\cite{abides21} is the reference market simulation configuration introduced in ABIDES-gym. It includes all agents in RFD, along with an additional percentage-of-volume~(POV) agent that provides extra liquidity to the simulated market.
    \item \textbf{LOBGAN}~\cite{coletta23} is a conditional Wasserstein GAN with gradient penalty, trained to generate the next order conditioned on market history.
\end{itemize}
    
\begin{table*}[!t]
  \addtolength\belowcaptionskip{-0.5cm}
\centering
\small
\begin{tabular}{@{}crrrr@{}}
\toprule
\textbf{Environment} & \textbf{Ret(\%)($\uparrow$)} & \textbf{Vol($\downarrow$)} & \textbf{SR($\uparrow$)} & \textbf{MDD(\%)($\downarrow$)} \\ \midrule
Replay & $0.009_{\pm 0.043}$ & $0.413_{\pm 0.090}$ & $0.014_{\pm 0.008}$ & $1.133_{\pm 0.173}$ \\
RFD & $0.000_{\pm 0.008}$ & $0.159_{\pm 0.094}$ & $0.011_{\pm 0.029}$ & $0.803_{\pm 0.327}$ \\
DigMA-c & $0.015_{\pm 0.023}$ & $\boldsymbol{0.147_{\pm 0.151}}$ & $0.006_{\pm 0.066}$ & $\boldsymbol{0.715_{\pm 0.464}}$ \\
DigMA & $\boldsymbol{0.029_{\pm 0.019}}$ & $0.411_{\pm 0.121}$ & $\boldsymbol{0.049_{\pm 0.031}}$ & $1.313_{\pm 0.156}$ \\ \bottomrule
\end{tabular}
\caption{Average out-of-sample test results (in percentage). Best results are highlighted with bold face.}
\label{tab:rl_task}
% }
\end{table*}

We assess fidelity by examining the distributional discrepancies of several canonical ``stylized facts,'' which are key statistical properties of asset returns and order-book dynamics, between real and simulated markets. These statistics capture widely studied characteristics of financial market microstructure:
\begin{itemize}
\item \textbf{Minutely Log Returns~(MinR)} are the logarithmic differences between consecutive minute-level prices.
\item \textbf{Return Auto-correlation~(RetAC)} is the linear auto-correlation between the return series and its lagged values. Empirical studies on real market data show that returns exhibit little to no auto-correlation at short lags.
\item \textbf{Volatility Clustering~(VolC)} is the linear auto-correlation of squared returns and their lagged values. It reflects the empirical observation that periods of high volatility tend to cluster over time.
\item \textbf{Order Imbalance Ratio~(OIR)} is the proportional volume difference between the best bid and best ask, capturing directional trading tendencies of market participants.
\end{itemize}
Additional details are provided in the appendix.

We report results using the unconditional variant of DigMA to isolate the effect of controllability from fidelity. Figure~\ref{fig:sfd} illustrates the distributions of these statistics, where the real-market distributions are shown as \textit{Real} using a solid golden line. Across all metrics, DigMA more closely matches the real distributions than competing methods.
To quantify these discrepancies, we compute the \textbf{Kullback--Leibler (K-L) divergence} between the real and simulated statistics.
Table~\ref{tab:dist} summarizes the results, demonstrating that DigMA achieves the lowest K-L divergence on most metrics.
While LOBGAN attains the lowest divergence on RetAC due to its auto-regressive order generation, DigMA substantially outperforms LOBGAN on the remaining stylized facts by a large margin.
Overall, these results indicate that DigMA achieves superior fidelity in market simulation, effectively capturing realistic market dynamics.

\subsection{Evaluation on High-Frequency Trading with Reinforcement Learning}
We evaluate the usefulness of DigMA as a training environment for reinforcement learning (RL) agents in a high-frequency trading task.

\subsubsection{Settings}
We train a trading agent in simulated market environments using the A2C algorithm, with the objective of optimizing high-frequency trading performance. 
The agent takes a discrete action every 10 seconds. 
The action space includes buying or selling at any of the best five price levels with an integer volume between 1 and 10 units, as well as an option to take no action. 
The observation space consists of price changes over the past twenty seconds, ten-level bid--ask price--volume pairs, and account information including capital, position, and cash.

Each agent is trained in environments generated by one of the following methods: historical replay, RFD, DigMA, and a variant of DigMA without meta-controller conditioning (DigMA-c). 
Each training run lasts 200 episodes, with each episode corresponding to a full trading day containing 1,440 decision steps over four trading hours.
After training, agents are evaluated for 50 episodes in an environment that replays out-of-sample real market data. 
This evaluation is repeated three times using three non-overlapping out-of-sample periods to assess robustness. 
All RL models are trained with identical hyperparameters to ensure fairness.

We evaluate the performance of the RL trading task using four metrics. 
\textbf{Daily return~(Ret)} is defined as the mean return across episodes and is used to assess profitability. 
\textbf{Daily volatility~(Vol)} is defined as the standard deviation of daily returns and is used to assess risk. 
\textbf{Sharpe ratio~(SR)} is defined as the ratio of daily return to volatility, reflecting the return--risk trade-off. 
\textbf{Maximum drawdown~(MDD)} is defined as the largest intraday decline in cumulative profit during testing and is used to assess robustness under extreme market conditions. 
For daily return and Sharpe ratio, higher values indicate better performance, whereas for daily volatility and maximum drawdown, lower values are preferred.

\subsubsection{Results}
Table~\ref{tab:rl_task} reports the average numerical results of the high-frequency trading task. The trading agent trained in the DigMA-generated environment achieves the highest daily return and Sharpe ratio among all baselines. The agent trained with DigMA-c attains the lowest daily volatility and maximum drawdown, indicating the emergence of a more conservative trading strategy. These results demonstrate that DigMA provides a more effective environment for policy learning and enables the trading agent to learn a better policy. Furthermore, the performance differences between DigMA and DigMA-c highlight that enabling controllability in the simulated training environment influences the decision-making preferences of RL trading agents.

\subsection{Analysis of Computational Efficiency}
In this section, we compare the computational efficiency of DigMA with all baseline models. We evaluate efficiency using the latency of financial market generation, defined as the average time required to generate a single order. 
As shown in~\Cref{tab:efficiency}, DigMA achieves the fastest order generation speed among all methods, requiring only approximately 0.017 milliseconds per order, which makes it suitable for real-time and latency-critical applications. RFD and RMSC exhibit slightly higher latency due to their designs, which require querying the virtual exchange at each decision step. LOBGAN is approximately 100 times slower than DigMA, owing to its recurrent neural network architecture and autoregressive generation process. Overall, DigMA provides the highest computational efficiency for financial market generation.

\begin{table}[!t]
  \addtolength\belowcaptionskip{-0.5cm}
\centering
\small
\begin{tabular}{@{}lrr@{}}
    \toprule
    \textbf{Model} & \textbf{Time(ms)/Order}  \\
    \midrule
    RFD & 0.049 \\
    RMSC & 0.075 \\
    LOBGAN & 1.710 \\
    DigMA & \textbf{0.017} \\
    \bottomrule
\end{tabular}
\caption{Comparison on computational efficiency.}
\label{tab:efficiency}
\end{table}

\section{Conclusion and Future Work}\label{sec:con}
In this paper, we formulate the problem of controllable financial market generation and propose the \textit{\underline{Di}ffusion \underline{G}uided \underline{M}eta \underline{A}gent}~(DigMA) model to address it. 
Specifically, we employ a diffusion model to capture the dynamics of market states, which are represented by time-evolving distribution parameters of mid-price return rates and order arrival rates, and we define a meta agent with financial economic priors to generate orders from the corresponding distributions. 
Extensive experimental results demonstrate that DigMA achieves superior controllability and generation fidelity. 
While this work focuses on generating the order flow of a single stock at each time step, future work will consider the correlations among multiple assets to generate more realistic markets.

\section{Acknowledgments}
This work is supported by NSFC Project~(No.62192783, No.12326615).
\bibliography{main}
\section*{Disclaimer}

The DigMA model is provided “as is”, without warranty of any kind, express or implied, including but not limited to the warranties of merchantability, fitness for a particular purpose and noninfringement. The DigMA model is aimed to facilitate research and development process in the financial industry and not ready-to-use for any financial investment or advice. Users shall independently assess and test the risks of the DigMA model in a specific use scenario, ensure the responsible use of AI technology, including but not limited to developing and integrating risk mitigation measures, and comply with all applicable laws and regulations in all applicable jurisdictions. The DigMA model does not provide financial opinions or reflect the opinions of Microsoft, nor is it designed to replace the role of qualified financial professionals in formulating, assessing, and approving finance products. The inputs and outputs of the DigMA model belong to the users and users shall assume all liability under any theory of liability, whether in contract, torts, regulatory, negligence, products liability, or otherwise, associated with use of the DigMA model and any inputs and outputs thereof.
\appendix
\clearpage
\section{Detailed Description of the Dataset and Preprocessing}\label{app:data}
We conduct experiments on two tick-by-tick order-book datasets from the China A-share market: \textit{A-Main} and \textit{ChiNext}. Both datasets are collected from Wind\footnote{https://www.wind.com.cn/. According to the data license, we are unable to share the full original data.} and cover the Shenzhen Stock Exchange (SZSE) for the year 2020. Table~\ref{tab:app_data} summarizes the dataset statistics.

\begin{table}[ht]
\centering
\small
\begin{tabular}{@{}lcc@{}}
\toprule
\textbf{Statistic} & \textbf{A-Main} & \textbf{ChiNext} \\ \midrule
Number of date--stock pairs & 316,287 & 122,574 \\
Number of unique stocks & 1,452 & 854 \\
Number of unique trading days & 237 & 231 \\ \bottomrule
\end{tabular}
\caption{Summary statistics of the DigMA datasets.}
\label{tab:app_data}
\end{table}

For each dataset, we randomly sample 5{,}000 instances for validation and 5{,}000 instances for testing, and use the remaining samples for training.

The preprocessing pipeline consists of filtering and transformation steps.
\begin{itemize}
    \item \textbf{Filtering.} We remove samples containing incomplete records (e.g., trading suspensions) or invalid orders (e.g., invalid order prices).
    
    \item \textbf{Transformation.} We transform the tick-by-tick data into market states represented by mid-price returns and order arrival rates.
    
    For the mid-price return, we first extract a minutely mid-price series from the order flow of each trading day. The mid-price at minute $t$ is defined as
    $p_t = (a_{t,1} + b_{t,1}) / 2$,
    where $a_{t,1}$ and $b_{t,1}$ denote the best bid and best ask prices at the end of the $t$-th trading minute, respectively. We then compute the log-difference between consecutive minutes to obtain the mid-price return
    $r_t = \log(p_t) - \log(p_{t-1})$.
    The resulting return sequence is denoted as $\boldsymbol{r} = [r_1, r_2, \ldots, r_T]$. The effective sequence length is $T = 236$, excluding the call auction phase at the end of each trading day.
    
    For the order arrival rate, we first compute the number of orders within each minute, denoted by $N_t$. Assuming that order arrivals follow a Poisson process, we treat $N_t$ as the expected number of arrivals and approximate the order arrival rate as $\lambda_t = N_t$. The resulting sequence is $\boldsymbol{\lambda} = [\lambda_1, \lambda_2, \ldots, \lambda_T]$.
    
    During training, all input features are normalized using z-score normalization, with the mean and standard deviation computed from the training split. During sampling, the model outputs are inverse-transformed accordingly.
\end{itemize}

\section{Detailed Description of DigMA Training}\label{app:ddpm}
\subsection{Brief Review of DDPM Model}
A diffusion probabilistic model~\cite{SD15} learns to reverse the transitions of a Markov chain which is known as the diffusion process that gradually adds noise to data, ultimately destroying the signal. 

Let $\mathbf{x}_0 \in \mathbb{R}^d \sim q(\mathbf{x}_0)$ denote a data sample of dimension $d$ from space $\mathcal{X}$. The forward diffusion process generates a sequence $\mathbf{x}_1, \ldots, \mathbf{x}_N$ with the same dimensionality as $\mathbf{x}_0$ by progressively adding Gaussian noise over $N$ time steps:
\begin{equation}
q(\mathbf{x}_1, \ldots, \mathbf{x}_N \mid \mathbf{x}_0) := \prod_{n=1}^{N} q(\mathbf{x}_n \mid \mathbf{x}_{n-1}).
\end{equation}
Each transition kernel is defined as
\begin{equation}
q(\mathbf{x}_n \mid \mathbf{x}_{n-1}) := \mathcal{N}\!\left(\mathbf{x}_n; \sqrt{1-\beta_n}\,\mathbf{x}_{n-1}, \beta_n \mathbf{I}\right),
\end{equation}
where $\{\beta_n \in (0,1)\}_{n=1}^N$ denotes a predefined variance schedule.

It follows that the marginal distribution of $\mathbf{x}_n$ admits a closed form:
\begin{equation}
q(\mathbf{x}_n \mid \mathbf{x}_0) = \mathcal{N}\!\left(\mathbf{x}_n; \sqrt{\bar{\alpha}_n}\,\mathbf{x}_0, (1-\bar{\alpha}_n)\mathbf{I}\right),
\end{equation}
where $\alpha_n := 1-\beta_n$ and $\bar{\alpha}_n := \prod_{s=1}^{n} \alpha_s$.

The reverse process is parameterized by $\theta$ and defined as
\begin{equation}
p_\theta(\mathbf{x}_0, \mathbf{x}_1, \ldots, \mathbf{x}_N)
:= p(\mathbf{x}_N)\prod_{n=1}^{N} p_\theta(\mathbf{x}_{n-1} \mid \mathbf{x}_n),
\end{equation}
where
\begin{equation}
p_\theta(\mathbf{x}_{n-1} \mid \mathbf{x}_n)
:= \mathcal{N}\!\left(\mathbf{x}_{n-1};
\boldsymbol{\mu}_\theta(\mathbf{x}_n,n),
\boldsymbol{\Sigma}_\theta(\mathbf{x}_n,n)\right),
\end{equation}
and the prior is given by $p(\mathbf{x}_N)=\mathcal{N}(\mathbf{0},\mathbf{I})$.

The training objective minimizes the negative log-likelihood,
\begin{equation}
L := \mathbb{E}\big[-\log p_\theta(\mathbf{x}_0)\big]
\leq
\mathbb{E}_q\!\left[-\log
\frac{p_\theta(\mathbf{x}_0,\ldots,\mathbf{x}_N)}
{q(\mathbf{x}_1,\ldots,\mathbf{x}_N \mid \mathbf{x}_0)}
\right],
\end{equation}
which leads to a tractable variational bound. A commonly adopted parameterization expresses the mean as
\begin{equation}
\boldsymbol{\mu}_\theta(\mathbf{x}_n,n)
= \frac{1}{\sqrt{\alpha_n}}
\left(
\mathbf{x}_n -
\frac{\beta_n}{\sqrt{1-\bar{\alpha}_n}}
\boldsymbol{\epsilon}_\theta(\mathbf{x}_n,n)
\right),
\end{equation}
where $\boldsymbol{\epsilon}_\theta$ is a neural network predicting the injected noise. This parameterization yields the simplified objective
\begin{equation}
L_{\text{simple}}
:= \mathbb{E}_{\mathbf{x}_0,\boldsymbol{\epsilon},n}
\Big[
\big\|
\boldsymbol{\epsilon} -
\boldsymbol{\epsilon}_\theta(
\sqrt{\bar{\alpha}_n}\mathbf{x}_0 +
\sqrt{1-\bar{\alpha}_n}\boldsymbol{\epsilon}, n)
\big\|^2
\Big].
\end{equation}

During sampling, $\mathbf{x}_{n-1}$ is generated as
\begin{equation}
\mathbf{x}_{n-1}
=
\frac{1}{\sqrt{\alpha_n}}
\left(
\mathbf{x}_n -
\frac{1-\alpha_n}{\sqrt{1-\bar{\alpha}_n}}
\boldsymbol{\epsilon}_\theta(\mathbf{x}_n,n)
\right)
+ \sigma_n \mathbf{z},
\end{equation}
where $\sigma_n = \sqrt{\beta_n}$ and $\mathbf{z} \sim \mathcal{N}(\mathbf{0},\mathbf{I})$~\cite{ho20}.

To obtain a conditional DDPM using classifier-free guidance~\cite{ho21}, the noise estimator is augmented to accept a condition embedding $\boldsymbol{\phi}(\mathbf{c})$, yielding
$\boldsymbol{\epsilon}_\theta(\mathbf{x}_n,n,\boldsymbol{\phi}(\mathbf{c}))$.
During training, the condition $\mathbf{c}$ is replaced by an unconditional token $\mathbf{c}_0$ with probability $p_{\text{uncond}}$, producing the unconditional prediction
$\boldsymbol{\epsilon}_\theta(\mathbf{x}_n,n)=
\boldsymbol{\epsilon}_\theta(\mathbf{x}_n,n,\mathbf{c}_0)$.

During sampling, a guidance scale $s$ controls the conditioning strength by modifying the noise prediction as
\begin{equation}
\tilde{\boldsymbol{\epsilon}}_{\theta,\phi}(\mathbf{x}_n,n,\mathbf{c})
=
(1-s)\boldsymbol{\epsilon}_\theta(\mathbf{x}_n,n)
+
s\boldsymbol{\epsilon}_\theta(\mathbf{x}_n,n,\boldsymbol{\phi}(\mathbf{c})).
\end{equation}
Both conditional and unconditional sampling can be accelerated using DDIM~\cite{ddim}.

\subsection{Algorithmic Procedure for Training the Meta Controller}
The pesudocode of training meta controller is provided in~\Cref{algo:train}.

\subsection{Detailed Training Parameters of the Meta Controller}\label{app:param}
The denoising model $\boldsymbol{\epsilon}_\theta$ used in the meta controller is adapted from~\cite{ho20}. The input to the denoising model has shape $(B, C, T)$, where $B$ denotes the batch size, $C$ is the number of channels, and $T$ is the number of valid trading minutes per day. In our setting, $C = 2$ and $T = 236$.

The denoising model adopts a U-Net architecture consisting of three downsampling blocks, one middle block, three upsampling blocks, and one output block. Each upsampling and downsampling block contains two ResNet blocks, one self-attention layer, and one upsampling or downsampling layer. Each ResNet block comprises two convolutional layers with kernel size 15, followed by SiLU activations, residual connections, and layer normalization. Each upsampling or downsampling layer uses a scaling factor of 2. The number of channels starts at 64 and is multiplied by a factor of 4 after each downsampling stage. The middle block consists of two ResNet blocks with a self-attention layer in between. The output block consists of an additional ResNet block followed by a $1\times1$ convolutional layer.

In addition, the conditioning embedding is produced by a fully connected network with two layers, each of dimension 64. Table~\ref{tab:app_para} summarizes the key architectural and training parameters described above.

Each model is trained on a single NVIDIA Tesla V100 GPU for 10 epochs, requiring approximately 2 hours on the A-Main dataset and 1 hour on the ChiNext dataset.

\begin{algorithm}[!t]  
\small
\SetAlgoLined  
\KwData{Order flow $\boldsymbol{O}$ processed into market states $\boldsymbol{x}$\;} 
\KwResult{Meta controller network parameters $\theta$}

\Repeat{the maximum number of epochs is reached}{
Sample $\boldsymbol{x}_0 \sim q(\boldsymbol{x})$\;
Compute target indicator $c=\mathcal{F}(\boldsymbol{x})$\;
With probability $p_{\text{uncond}}$, replace $c$ with the unconditional identifier $c_u$\;
Sample time step $n \sim \mathcal{U}(\{1,\ldots,N\})$\;
Sample noise $\boldsymbol{\epsilon} \sim \mathcal{N}(\boldsymbol{0},\boldsymbol{I})$\;
Corrupt the data:
$\boldsymbol{x}_n = \sqrt{\bar{\alpha}_n}\boldsymbol{x}_0 + \sqrt{1-\bar{\alpha}_n}\boldsymbol{\epsilon}$\;
Update parameters by gradient descent on
\[
    \nabla_{\theta,\phi}
    \left\|
    \boldsymbol{\epsilon} -
    \tilde{\boldsymbol{\epsilon}}_{\theta,\phi}(\boldsymbol{x}_n, n, c)
    \right\|^2;
    \]
}
\caption{Training procedure of the DigMA meta controller} 
\label{algo:train}
\end{algorithm}

\section{Details of Generation with DigMA}

\begin{table}[!t]
\centering
\small
\begin{tabular}{@{}cc@{}}
\toprule
\textbf{Parameter} & \textbf{Value} \\ \midrule
Denoising input shape $(C, T)$ & $(2, 236)$ \\
Number of diffusion steps & 200 \\
Number of ResNet blocks per stage & 2 \\
Base channel dimension & 64 \\
U-Net channel multipliers & $(1, 4, 16)$ \\
Embedding dimension & 256 \\
Convolution kernel size & 15 \\
Convolution padding & 7 \\
Unconditional probability $p_{\text{uncond}}$ & 0.5 \\
Batch size & 256 \\
Learning rate & $10^{-5}$ \\ 
\bottomrule
\end{tabular}
\caption{Architectural and training hyperparameters for the meta controller.}
\label{tab:app_para}
\end{table}

\subsection{Algorithmic Procedure for Generating Financial Markets with DigMA}
The pseudocode of the generation algorithm is provided in~\Cref{app:genalgo}. 

\begin{algorithm*}[t]  
\small
\SetAlgoLined  
\KwIn{Meta controller parameters $\theta$, control target $c$, guidance scale $s$, maximum trading time $T$, initial price $p_0$, meta agent parameters $\lambda_f, \lambda_c, \lambda_n, \tau_0, \alpha_0, \sigma_n$, simulated exchange}
\KwOut{Order flow $\boldsymbol{O}$}  
\textbf{Phase 1: Sampling market states with the meta-controller}\\
Sample $\boldsymbol{x}_N \sim \mathcal{N}(\boldsymbol{0},\boldsymbol{I})$\;
\For{$n = N$ \KwTo $1$}{
    Sample $\boldsymbol{z} \sim \mathcal{N}(\boldsymbol{0},\boldsymbol{I})$\;
    $\hat{\boldsymbol{\epsilon}} \gets
    \tilde{\boldsymbol{\epsilon}}_{\theta,\phi}(\boldsymbol{x}_n, n, c)
    =
    (1-s)\boldsymbol{\epsilon}_\theta(\boldsymbol{x}_n,n)
    + s\,\boldsymbol{\epsilon}_\theta(\boldsymbol{x}_n,n,\boldsymbol{\phi}(c))$\;
    $\boldsymbol{x}_{n-1} \gets
    \frac{1}{\sqrt{\alpha_n}}
    \left(
    \boldsymbol{x}_n -
    \frac{1-\alpha_n}{\sqrt{1-\bar{\alpha}_n}}
    \hat{\boldsymbol{\epsilon}}
    \right)
    + \sigma_n \boldsymbol{z}$\;
}

\textbf{Phase 2: Generating order flow with the meta-agent}\\
Initialize $t = 0$, $\boldsymbol{O} = \emptyset$, $p_t = p_0$\;
\Repeat{$t < T$}{
    Extract $r_t$ and $\lambda_t$ from $\boldsymbol{x}$\;
    Initialize agent $\mathcal{A}_i$:
    $S_i \sim \operatorname{Exponential}(S_0)$,
    $C_i \sim \operatorname{Exponential}(C_0)$\;
    Sample heterogeneity factors:
    $g_f \sim \operatorname{Laplace}(\lambda_f)$,
    $g_c \sim \operatorname{Laplace}(\lambda_c)$,
    $g_n \sim \operatorname{Laplace}(\lambda_n)$\;
    Set
    $\tau_i = \tau_0 \frac{1+g_f}{1+g_c}$,
    $\alpha_i = \alpha_0 \frac{1+g_f}{1+g_c}$\;
    Sample inter-arrival time $\delta_i \sim \operatorname{Exponential}(\lambda_t)$,
    set $t_i = t + \delta_i$\;
    Observe market return $\bar{r}$ and sample noise
    $r_\sigma \sim \mathcal{N}(0,\sigma_n)$\;
    Estimate future return
    $\hat{r} =
    \frac{g_f r_t + g_c \bar{r} + g_n r_\sigma}{g_f + g_c + g_n}$\;
    Estimate future price $\hat{p}_t = p_t \exp(\hat{r})$\;
    Solve $p_l$ from
    $p_l (u(p_l) - S_i) = C_i$,
    where
    $u(p) = \frac{\ln(\hat{p}_t / p)}{\alpha_i V p}$\;
    Sample order price $p_i \sim \mathcal{U}(p_l, \hat{p}_t)$\;
    Compute order size $q_i = u(p_i) - S_i$\;
    Determine order type $o_i = \operatorname{sign}(q_i)$\;
    Form order $\boldsymbol{o}_i = (t_i, p_i, q_i, o_i)$\;
    Update price
    $p_t \gets \operatorname{Exchange}(\boldsymbol{o}_i)$\;
    Update $t \gets t_i$,
    $\boldsymbol{O} \gets \boldsymbol{O} \cup \{\boldsymbol{o}_i\}$\;
}

\caption{Generating market order flow with DigMA}  
\label{app:genalgo}
\end{algorithm*}  

\subsection{Discussion on Meta Controller Sampling Parameters}\label{app:parm_controller}
During sampling with the meta controller, we employ DDIM~\cite{ddim} to reduce the number of sampling steps to 20. To obtain high-quality generation results, the classifier guidance scale $s$ must be carefully selected. In our experiments, we choose the optimal value of $s$ for each target scenario from the set $\{1, 2, 4, 6, 8\}$. The selection criterion is based on the discrepancy between the control target and the generated scenario evaluated on the validation set, and we adopt the value of $s$ that yields the lowest discrepancy.

\subsection{Discussion on Meta Agent Parameters}\label{app:parm_agent}
Following the heterogeneous agent framework in~\cite{chiarella09}, the meta agent incorporates several probabilistic parameters to model agent heterogeneity. Specifically, $\tau_0$ denotes the estimation horizon, $\alpha_0$ controls risk aversion, and $\sigma_n$ represents the noise level in returns. These parameters are designed to capture heterogeneous human preferences observed in real-world stock markets.

In all experiments, we fix $\tau_0 = 30$, $\alpha_0 = 0.1$, $\sigma_n = 10^{-4}$, and $p_0 = 10$ across all runs to avoid overfitting these parameters. Nevertheless, they can be calibrated to further improve the fidelity of the generated market dynamics.

\subsection{Definition of Stylized Facts and Implementation Details}

We assess the fidelity of simulated markets by comparing the distributions of several canonical \emph{stylized facts} between real and generated order flow. Stylized facts are well-established statistical regularities of asset returns and order-book dynamics, and matching these properties is widely regarded as a key criterion for evaluating the realism of synthetic financial markets.

Let $\{p_t\}_{t=0}^{T}$ denote the sequence of minute-level transaction prices. Based on this price series and the corresponding limit order book snapshots, we compute the following stylized facts.

\subsection{Minutely Log Returns (MinR)}
The minutely log return is defined as the logarithmic difference between consecutive minute-level prices:
\begin{equation}
r_t = \log p_t - \log p_{t-1}, \quad t = 1, \dots, T.
\end{equation}
The empirical distribution of $\{r_t\}$ characterizes short-horizon return behavior, including tail properties and asymmetry.

\subsubsection{Return Auto-correlation (RetAC)}
Return auto-correlation measures the linear dependence between consecutive returns. In our evaluation, we focus on the first-order (minimum-lag) auto-correlation, defined as
\begin{equation}
\mathrm{RetAC} = \mathrm{Corr}(r_t, r_{t-1}),
\end{equation}
where $\mathrm{Corr}(\cdot,\cdot)$ denotes the Pearson correlation coefficient.
Empirical studies consistently show that asset returns exhibit little to no auto-correlation at short lags, making this statistic a sensitive indicator of market realism.

\begin{table*}[!t]
\centering
\setlength{\tabcolsep}{3pt}
\newcolumntype{R}{>{\raggedleft\arraybackslash\hspace*{\fill}}p{1.1cm}<{\hspace{0.2cm}\strut}}

\small
\label{tab:addt1}
\begin{tabular}{@{}cccRRRRRRRRRR@{}}
\toprule
 &  &  & \multicolumn{5}{c}{\textbf{A-Main}} & \multicolumn{5}{c}{\textbf{ChiNext}} \\ 
 \cmidrule(lr){4-8}\cmidrule(lr){9-13}
\textbf{Target} & \textbf{Method} & & \multicolumn{1}{c}{\textbf{Lower}} & \multicolumn{1}{c}{\textbf{Low}} & \multicolumn{1}{c}{\textbf{Medium}} & \multicolumn{1}{c}{\textbf{High}} & \multicolumn{1}{c}{\textbf{Higher}} & \multicolumn{1}{c}{\textbf{Lower}} & \multicolumn{1}{c}{\textbf{Low}} & \multicolumn{1}{c}{\textbf{Medium}} & \multicolumn{1}{c}{\textbf{High}} & \multicolumn{1}{c}{\textbf{Higher}} \\ \midrule
\multirow{6}{*}{Return} & \multirow{2}{*}{No Control} & mean & 1.443 & 0.583 & 0.529 & 0.813 & 2.337 & 0.979 & 0.684 & 0.992 & 1.718 & 3.923 \\
 &  & std & 0.211 & 0.096 & 0.048 & 0.072 & 0.201 & 0.134 & 0.095 & 0.103 & 0.133 & 0.209 \\
 & \multirow{2}{*}{Discrete} & mean & 1.055 & 0.494 & 0.228 & 0.429 & 0.664 & 1.285 & 0.807 & 0.243 & \textbf{0.413} & 0.869 \\
 &  & std & 0.159 & 0.040 & 0.003 & 0.027 & 0.031 & 0.008 & 0.055 & 0.023 & 0.029 & 0.110 \\
 & \multirow{2}{*}{Continuous} & mean & \textbf{0.206} & \textbf{0.178} & \textbf{0.161} & \textbf{0.184} & \textbf{0.212} & \textbf{0.584} & \textbf{0.539} & \textbf{0.342} & 0.449 & \textbf{0.840} \\
 &  & std & 0.080 & 0.023 & 0.022 & 0.017 & 0.033 & 0.211 & 0.267 & 0.389 & 0.422 & 0.522 \\ \midrule
\multirow{6}{*}{Amplitude} & \multirow{2}{*}{No Control} & mean & 0.521 & 0.268 & 0.268 & 0.699 & 3.298 & 1.130 & 0.638 & 0.427 & 0.608 & 2.763 \\
 &  & std & 0.071 & 0.044 & 0.017 & 0.024 & 0.105 & 0.074 & 0.077 & 0.081 & 0.088 & 0.107 \\
 & \multirow{2}{*}{Discrete} & mean & \textbf{0.049} & 0.088 & 0.309 & 0.502 & 0.930 & \textbf{0.057} & 0.134 & 0.346 & 0.523 & \textbf{0.963} \\
 &  & std & 0.005 & 0.004 & 0.016 & 0.016 & 0.053 & 0.003 & 0.008 & 0.002 & 0.031 & 0.015 \\
 & \multirow{2}{*}{Continuous} & mean & 0.054 & \textbf{0.076} & \textbf{0.149} & \textbf{0.247} & \textbf{0.348} & 0.110 & \textbf{0.116} & \textbf{0.255} & \textbf{0.437} & 0.973 \\
 &  & std & 0.017 & 0.016 & 0.017 & 0.080 & 0.011 & 0.007 & 0.011 & 0.045 & 0.065 & 0.113 \\ \midrule
\multirow{6}{*}{Volatility} & \multirow{2}{*}{No Control} & mean & 0.021 & 0.115 & 0.431 & 1.209 & 4.288 & 0.029 & 0.246 & 0.713 & 1.737 & 5.221 \\
 &  & std & 0.003 & 0.018 & 0.038 & 0.065 & 0.123 & 0.006 & 0.015 & 0.034 & 0.061 & 0.116 \\
 & \multirow{2}{*}{Discrete} & mean & 0.016 & 0.123 & 0.383 & 0.890 & 2.393 & 0.029 & 0.188 & 0.481 & \textbf{0.948} & \textbf{2.257} \\
 &  & std & 0.001 & 0.017 & 0.008 & 0.021 & 0.008 & 0.003 & 0.007 & 0.014 & 0.016 & 0.019 \\
 & \multirow{2}{*}{Continuous} & mean & \textbf{0.011} & \textbf{0.104} & \textbf{0.318} & \textbf{0.774} & \textbf{2.389} & \textbf{0.028} & \textbf{0.178} & \textbf{0.473} & 1.016 & 2.631 \\
 &  & std & 0.001 & 0.010 & 0.030 & 0.057 & 0.098 & 0.005 & 0.008 & 0.020 & 0.078 & 0.260 \\ \bottomrule
\end{tabular}
\caption{Mean squared error (MSE) between the target indicators and the aggregated statistics of the generated order flow. Results are reported as mean and standard deviation over three independent runs. Best performance is highlighted in bold.}
\end{table*}

\begin{table*}[!t]
\centering
\small
\label{tab:add_t2}
\begin{tabular}{@{}lcrrrrrrrr@{}}
\toprule
 &  & \multicolumn{4}{c}{\textbf{A-Main}} & \multicolumn{4}{c}{\textbf{ChiNext}} \\ 
 \cmidrule(lr){3-6}\cmidrule(lr){7-10}
\textbf{Model} &  & \textbf{MinR} & \textbf{RetAC} & \textbf{VolC} & \textbf{OIR} & \textbf{MinR} & \textbf{RetAC} & \textbf{VolC} & \textbf{OIR} \\ 
\midrule
\multirow{2}{*}{RFD} & mean & 1.198 & 5.010 & 0.839 & 0.015 & 0.272 & 2.987 & 0.691 & 0.022 \\
 & std & 0.083 & 1.335 & 0.322 & 0.001 & 0.037 & 0.937 & 0.316 & 0.001 \\
\multirow{2}{*}{RMSC} & mean & 2.640 & 10.170 & 1.237 & 0.563 & 1.371 & 7.461 & 0.668 & 0.588 \\
 & std & 0.051 & 0.204 & 0.959 & 0.016 & 0.001 & 0.195 & 0.243 & 0.016 \\
\multirow{2}{*}{LOBGAN} & mean & 0.151 & \textbf{1.903} & 1.101 & 0.309 & 0.135 & \textbf{1.711} & 0.507 & 0.282 \\
 & std & 0.007 & 1.045 & 0.639 & 0.011 & 0.002 & 1.043 & 0.108 & 0.010 \\
\multirow{2}{*}{DigMA} & mean & \textbf{0.084} & 2.781 & \textbf{0.273} & \textbf{0.009} & \textbf{0.079} & 1.997 & \textbf{0.218} & \textbf{0.009} \\
 & std & 0.006 & 0.417 & 0.020 & 0.001 & 0.002 & 0.243 & 0.049 & 0.001 \\ \bottomrule
\end{tabular}
\caption{Kullback--Leibler (KL) divergence between the distributions of stylized facts computed from real and simulated order flow. Results are reported as mean and standard deviation over three independent runs. Lower values indicate better agreement.}
\end{table*}

\subsection{Volatility Clustering (VolC)}
Volatility clustering captures the tendency of large price changes to be followed by large changes, and small changes by small ones. We quantify volatility clustering using the first-order auto-correlation of squared returns:
\begin{equation}
\mathrm{VolC} = \mathrm{Corr}(r_t^2, r_{t-1}^2).
\end{equation}
A positive and persistent value of $\mathrm{VolC}$ is a hallmark of realistic financial time series. In practice, we compute this statistic using minute-level returns and their one-step lagged counterparts.

\subsection{Order Imbalance Ratio (OIR)}
Order imbalance ratio quantifies directional pressure in the limit order book by comparing bid-side and ask-side liquidity at the best price level. At time $t$, it is defined as
\begin{equation}
\mathrm{OIR}_t =
\frac{V^{\mathrm{bid}}_{t,1} - V^{\mathrm{ask}}_{t,1}}
     {V^{\mathrm{bid}}_{t,1} + V^{\mathrm{ask}}_{t,1}},
\end{equation}
where $V^{\mathrm{bid}}_{t,1}$ and $V^{\mathrm{ask}}_{t,1}$ denote the volumes at the best bid and best ask (level-1) prices, respectively.
The resulting imbalance values are aggregated to the minute level by retaining the last observed value within each minute.

All stylized facts are computed using identical definitions and aggregation procedures for both real and simulated markets, ensuring that any observed discrepancy reflects genuine differences in market dynamics rather than implementation artifacts.

\section{Detailed Definitions of Evaluation Metrics for the Trading Task}

We evaluate the performance of the reinforcement learning (RL) trading task using four standard investment metrics that jointly capture profitability, risk, and robustness. All metrics are computed based on the agent’s capital trajectory during the out-of-sample testing period.

Let $C_t$ denote the agent’s total capital at the end of day $t$. The daily return is computed as
\begin{equation}
r_t = \frac{C_t}{C_{t-1}} - 1.
\end{equation}

\paragraph{Daily Return (Ret).}
Daily return is defined as the mean of daily returns across evaluation episodes:
\begin{equation}
\mathrm{Ret} = \mathbb{E}[r_t],
\end{equation}
and is used to measure the overall profitability of the trading strategy.

\paragraph{Daily Volatility (Vol).}
Daily volatility is defined as the standard deviation of daily returns:
\begin{equation}
\mathrm{Vol} = \sqrt{\mathrm{Var}(r_t)},
\end{equation}
which quantifies the variability of returns and serves as a measure of risk.

\paragraph{Sharpe Ratio (SR).}
The Sharpe ratio evaluates the trade-off between return and risk, and is defined as
\begin{equation}
\mathrm{SR} = \frac{\mathbb{E}[r_t]}{\sqrt{\mathrm{Var}(r_t)}},
\end{equation}
where a higher value indicates better risk-adjusted performance.

\begin{table*}[!t]
  \addtolength\belowcaptionskip{-0.3cm}
\centering
\small
\begin{tabular}{@{}ccrrrr@{}}
\toprule
\textbf{Environment} & \textbf{Period} & \textbf{Ret(\%)($\uparrow$)} & \textbf{Vol($\downarrow$)} & \textbf{SR($\uparrow$)} & \textbf{MDD(\%)($\downarrow$)} \\ \midrule
\multirow{4}{*}{Replay} & 1 & 0.031 & 0.502 & 0.012 & 1.198 \\
 & 2 & 0.036 & 0.323 & 0.023 & 0.936 \\
 & 3 & -0.040 & 0.413 & 0.007 & 1.264 \\ \cmidrule(l){2-6} 
 & Average & $0.009_{\pm 0.043}$ & $0.413_{\pm 0.090}$ & $0.014_{\pm 0.008}$& $1.133_{\pm 0.173}$ \\ \midrule
\multirow{4}{*}{RFD} & 1 & -0.004 & 0.265 & -0.012 & 1.004 \\
 & 2 & 0.009 & 0.084 & 0.044 & 0.426 \\
 & 3 & -0.005 & 0.128 & 0.001 & 0.980 \\ \cmidrule(l){2-6} 
 & Average & $0.000_{\pm 0.008}$ & $0.159_{\pm 0.094}$ & $0.011_{\pm 0.029}$ & $0.803_{\pm 0.327}$ \\ \midrule
\multirow{4}{*}{DigMA-c} & 1 & 0.037 & 0.320 & 0.025 & 1.250 \\
 & 2 & 0.017 & 0.074 & 0.060 & 0.472 \\
 & 3 & -0.009 & 0.046 & -0.068 & 0.422 \\ \cmidrule(l){2-6} 
 & Average & $0.015_{\pm 0.023}$ & $\boldsymbol{0.147_{\pm 0.151}}$ & $0.006_{\pm 0.066}$ & $\boldsymbol{0.715_{\pm 0.464}}$ \\ \midrule
\multirow{4}{*}{DigMA} & 1 & 0.033 & 0.550 & 0.023 & 1.488 \\
 & 2 & 0.046 & 0.330 & 0.084 & 1.188 \\
 & 3 & 0.009 & 0.352 & 0.040 & 1.264 \\ \cmidrule(l){2-6} 
 & Average & $\boldsymbol{0.029_{\pm 0.019}}$ & $0.411_{\pm 0.121}$ & $\boldsymbol{0.049_{\pm 0.031}}$ & $1.313_{\pm 0.156}$ \\ \bottomrule
\end{tabular}
\caption{Detailed out-of-sample trading performance (percentages). Results are reported as mean $\pm$ standard deviation over three independent periods. Arrows indicate whether higher or lower values are preferred. Best results are highlighted in bold.}
\label{tab:rl_extra}
\end{table*}

\paragraph{Maximum Drawdown (MDD).}
Maximum drawdown measures the largest peak-to-trough decline of the capital process during the testing period. Formally, it is defined as
\begin{equation}
\mathrm{MDD}
=
\max_{t}
\left(
\max_{s \le t} C_s - C_t
\right),
\end{equation}
which captures the worst cumulative loss from a historical peak and is commonly used to assess robustness under adverse market conditions.

For all reported results, higher values of Ret and SR indicate better performance, whereas lower values of Vol and MDD are preferred.

\section{Additional Results}\label{app:std}
We report the complete result tables, including both the mean and standard deviation over three independent runs with different random seeds, for all main experiments. These results further confirm the statistical significance of DigMA’s performance advantages.

\end{document}